# Mathematize urbes by humanizing them: cities as isobenefit landscapes. Psycho-economical distances and personal isobenefit lines

Luca D'Acci

Highlights

- A city is viewed from different private visions: personal isobenefit lines visualize it.
- They transform aggregative habits to personal and isotropic spaces to anisotropic.
- Psycho-economical distances include pleasantness to cost, distance and time.

Abstract

The city reading proposed is a modern-postmodern urbanism approach which quantifies but by passing through subjectivism. The isobenefit lines shown translate cities into benefit landscapes, subjective and continually changeable according to personal moods/needs/preferences and urban transformations. They read attractiveness and how they flow throughout the city. Doing it for each urban point and for each urban attraction, we obtain the isobenefit orography of the city, namely a map of its urban attractions and of their flows. This is a liquid surface rather than solid, as it varies across time and people. It is in this liquidness where resides the complexity of cities, their bottom-up spirit and the dynamicity of equilibriums and networks. People do not necessarily go in the most accessible points, but where they need and want to, and, they flow through paths they need or choose to pass through. It is also introduced the likeability of places and paths: in addition to the usual parameters currently used – which weight distances in terms of physical distance, cost, time or mental easiness representations – psycho-economical distances used in the isobenefit lines proposed here, also consider how a place and a path pleases us. According to the Underground Hedonic Theory, this pleasure to pass through or to stay in agreeable areas has an underground and an inertia effect too which contributes to delight our lives. The final purpose of the science of cities and urban design is to understand cities and make them efficient and attractive to please our lives in them.

# 1. Introduction

Individual personal visions of their own cities may slightly or greatly differ among people. Formulizing this variety implies to consider a certain number of contemporary factors which influence personal views. A main aim of psycho-economical distances appearing in this paper is to extend the isobenefit lines (MIT Technology Review, 2012) into a heterogeneity and bottom-up pattern, where decision-making processes of each city dweller give their influence on the emergency of the complex system for antonomasia which is the city.

Following the romantic reaction to modernism, which began in the 1960s, against "the abstract platonic structures" of the modern universalism (Ellin, 1999), isobenefit lines, personalized by psycho-economical distances and individual preference criteria, respond also to pluralism and multiculturalism which are more and more characterizing our cities. Quoting Lynch, "cities are too complicated, too far beyond our control, and affect too many people, who are subject to too many cultural variations, to permit any rational answer. […] Someone might say 'I like Boston', but we all understand that this is merely a trivial preference, based on personal experience" (Lynch, 1984).

While isobenefit lines refer to the criteria of the majority of citizens of a city, where "majority" means the "ordinary" citizen (if this "ordinary" citizen exists; mainly if the variance of the preference criteria/behaviors of the "ordinary citizens" is limited), personal isobenefit lines refer to the criteria of each individual. Since the beginning of the postmodern urbanism reaction, urban sociologists began criticizing the environmental determinism of urban designers who do not consider how people perceive places. The consequent new field of environmental psychology underlined the "individual's personal identity in relation to the physical world through memories, ideas, feelings, attitudes, values, preferences, meanings, and conceptions about behavior relevant to the physical settings in his or her daily life" (Proshansky, 1990).

In this view, personal isobenefit lines humanize cities by transforming them as texts with many readings. As isobenefit lines read cities as 3 dimensional solids whose shapes diverge among cities and, for a same city, throughout timelines, personal isobenefit lines read cities as 3 dimensional solids whose shapes diverge among each person even for a same city in a same moment, and among different moods, times of life of a person.

This also in part fits into the shift from *complicated* to *complex* systems of the last second half of the past century.

In the 20s the system theory approach was dominant and suggested, during all the 50s, that systems were regarded as being centrally ordered, as a hierarchical sum of subsystems dominated by negative feedback, which implied a predominant controlled equilibrium status. Examples of these systems were also cities and regions. However, cities are never in equilibrium, they are constantly changing and dominated by positive feedback, not by negative's (Batty, 2012). A standard theory of cities was developed until the middle of the 20th century as an economic and transportation model based mostly on the monocentric city. Ideas and models were built on statistical aggregations of units, as for example models based on macro economics (econometric models, population models, Keynesian models).

In the 1970s (actually even earlier: "It was not Galileo or even Newton but Darwin that split this top-down world", Batty & Marshall, 2012), the idea changed: city was observed as controlled by positive feedback and not anymore from the top-down but from the bottom-up.

"[…] models were derived from work in a sub-area of artificial intelligence called distributed artificial intelligence (DAI). DAI aimed to solve problems by dividing them amongst a number of programs or agents, each with its own particular type of knowledge or expertise. In combination, the collection of agents would be better at finding solutions than any one agent working on its own. While DAI is primarily concerned with engineering effective solutions to real world problems, it was soon noticed that the technology of interacting intelligent agents could be applied to modelling social phenomena, with each agent representing one individual or organisational actor." (Gilbert & Terna, 2000).

A single agent may be able to reconfigure a complex system (system that have the potential to reconfigure themselves in ways that may be surprising, Batty & Torrneds, 2005), but the potential still exists for the system to change without us knowing the actions of any particular agent (Batty, 2012). Models were specified in more detail as, for example, by disaggregating into several types of populations, types of personal habits, etcetera. Fundamental elements themselves are to be represented: the so known agents.

The "new generation of thinking, based not on aggregative, equilibrium-seeking assumptions, consistent with models of how activities produce emergent social structures from the bottom up" (Epstein & Axtell, 1996), lies with a "new forms of representation at a fine spatial scale, in which units of space are conceived as cells and populations as individual agents, are currently changing the way we are able to simulate the evolution of cities" (Batty, 2005).

Models based on multi-agent decisions are becoming the dominant paradigm in any social simulation, due primarily to an agent-based worldview suggesting that complex systems emerge from the *bottom-up*, are highly decentralized, and are composed of a multitude of heterogeneous objects called agents ( Crooks, Castle, & Batty, 2008).

"Urban and regional modelling is a part of the broader and now fashionable field of complexity science […] there is a history of 50 years or more of serious development and therefore a substantial body of literature and ideas" (Wilson, 2012).

Including interactions among isobenefit lines, they show similarities with potential models, spatial interaction models, and more generally, with retail and gravitational models. They also work inside spatial equilibrium and location models: for a State-of-the-Art in Residential Location Models see, i.e., Pagliara, Preston, & Simmonds (2010), while for a Spatial Equilibrium reading, D'Acci, 2013a, D'Acci, 2013b and Glaeser, 2008.

More technically speaking, the methodology proposed in this paper can profitably be inserted in the wide framework in between GIS, Space Syntax, Urban Network Analysis and Multi agent based models (Batty, 2013).

If we consider this paper from the point of view of the change of urban attractiveness and of the relative frame origins-destinations of urban movements, during different times in the day, different days of the week, and different chosen paths, we could also refer, in some senses, to the Lund group's work on space-time prisms and volumes of the 1960s/70s, and more recent works proposed by Dykes, MacEachren, & Kraak (2005), Kraak (2003), Kraak and Ormeling (2011), Kwan and Neutens (2014), Mennis (2003), Miller (2005), and many others.

"Human activities interact and intertwine to create a complex social system that fulfills our physiological, economic, and social needs […] Hägerstrand's time geography offers a useful framework for studying individual activity and travel patterns under various constraints in a space–time context (Hägerstrand, 1970 and Hägerstrand, 1978, 1989)." (Shaw & Liu, 2009, p. 141).

The time geography framework helps the understanding of human spatial behavior, and the improvement of computational representations of the last decade has stimulated time geographic entities such as the space–time path and prism (Forer, 1998, Kuijpers et al., 2011, Miller, 1991, Miller, 1999 and Miller, 2005).

Examples of space–time GIS representing and analyzing individual activities and their interactions based on space–time paths and space–time prisms are available in Shaw and Liu (2009), Shaw, Yu, & Bombom, 2008, Yu (2006), Yu and Shaw (2008), while a semantic GIS aims to investigate "how humans represent geographic information in their minds" is presented in Mennis (2003).

Accessibility metrics typically use the shortest paths, in time or distance or cost, i.e., reach, gravity, betweeness, closeness, straightness (Sevtsuk, 2014).

Isobenefit lines analysis defines the concept of psycho-economical distances which add to the physical distances within locations the following elements: how fast, cheap and *pleasant* it is to move among locations. In other words, in addition to methods currently used in space syntax and to analyze complex networks in cities ( Batty, 2013, chaps. 6 and 7), psycho-economical distances say not just how fast, or cheap, or mentally easier it is to move among locations, but also how pleasant it is: you may choose one path instead of another not just because it is faster, cheaper or mentally easier, but because you *like* it more.

## 2. Isobenefit lines and psycho-economical distances

Isobenefit lines join the urban points with equal levels of benefit given from urban amenities. Depending from the aim of the analysis, these amenities may be schools, hospitals, libraries, or parks, pedestrian streets, nice squares, pleasant shopping areas, work places, etcetera.

The *Distributed Benefit* of a point *k* received from an amenity *i* distant *d*, and with a level *A* of attractiveness, is given by:

$$B_{i,k} = \frac{A_i}{1 + (d_{i-k}/E)}$$

(1)

where A is the *Punctual Benefit*: the benefit you receive when you are actually using the amenity without considering how difficult or expensive or uncomfortable it is to reach it. The line which graphically joins $A_i$ with $B_{i,k}$ is called *flow line* ( Fig. 2, Fig. 3, Fig. 4 and Fig. 15; it is as an electrical conductor which loses charge along the way because of several kinds of dispersion and resistance).
E is the variable which allows to transform the distances into *Psycho-Economical* distances.

$$E = \varepsilon \cdot E_{i-k}$$

(2)

$$E_{i-k} = \alpha P_{i-k} + \beta C_{i-k} + \gamma W_{i-k} + \delta Bi_{i-k}$$

(3)

Calculating equation 1 for all the *n* amenities present in the city:

$$B_k = \sum_{i=1}^{n} B_{i,k}$$

(4)

$E_{i-k}$ symbolizes the range of possibilities that the city presents for moving from *i* to *k* and *how* citizens move throughout the city: how much they use cars (C), public transport (P), bikes (Bi) and walking (W). $\varepsilon$ says *how much* they are willing to move around; it is a sort of subjective comfort of moving. In a certain way, $\varepsilon$ 'weighs' the advantage to enjoy numerous amenities (*Variety Value*), against the advantage of the proximity of one amenity (*Proximity Value*). If it is major than 1, it emphasizes the variety value; if minus, the proximity value.
The parameters $\alpha$, $\beta$, $\gamma$, $\delta$ describe how the citizen moves, what method he/she uses and in which relative percentage (each of them is equal or major than zero, and their sum must be 1). Under a certain distance from $A_i$, just $\gamma$ and $\delta$ count; after a certain distance they will count less and less.
These variables are quantified for each urban point *k* of the urban matrix, in relation with each amenity *i*, and are all translated into a same scale.
$P_{i-k}$ values the public transport system between *i* and *k*, and it is a weighted sum taking into account other under-variables (p) such as number of lines ($p_1$) and their frequency ($p_2$), speed ($p_3$), traffic ($p_4$), cost ($p_5$), distance of the stops from *i* and from *k* ($p_6$), objective comfort such as number of seating areas ($p_7$), noise ($p_8$), design ($p_9$). $C_{i-k}$ values the car transport system and it is a weighted sum of under-variables such as traffic ($c_1$) and costs ($c_2$) (oil, parking, car taxes, insurance, etc.). $W_{i-k}$ quantifies the facilities for pedestrians and includes the objective condition and availability of sidewalks and the quality of the streets from the pedestrian point of view (noise, esthetics, pollution, physical-psychological obstacles, etcetera). $Bi_{i-k}$ quantifies the same but from the point of view of biking.
The parameters which weight these under-variables composing the variables *P*, *C*, *W* and *Bi*, define the subjectivity and heterogeneity of citizens and informs how each under-variable is important for a specific dweller.
We call $v_E$ the vector of parameters $\alpha$, $\beta$, $\gamma$, $\delta$; $v_P$ the vector of the *m* parameters ($x_P$) of the *m* under-variables which personalize the variable *P*; $v_C$ the vector of *r* parameters ($x_C$) which personalize the variable *C*, $v_W$ the vector of the *t* parameters (*xw*) which personalize the variable *W*, and *vBi* the vector of *q* parameters (*xb*) which personalize the variable *Bi*.

$$P = x_{p_1}p_1 + x_{p_2}p_2 + ... + x_{pz}p_z + ... + x_{pm}p_m; \qquad v_p = (x_{p1}, x_{p2}, ..., x_{pz}, ..., x_{pm})$$
$$C = x_{1c}c_1 + ... + x_{cr}c_r; \qquad v_c = (x_{c1}, ..., x_{cr})$$
$$W = x_{w1}w_1 + ... + x_{wt}w_t; \qquad v_w = (x_{w1}, ..., x_{wt})$$
$$Bi = x_{bi1}bi_1 + ... + x_{biq}bi_q; \qquad v_{bi} = (x_{bi1}, ..., x_{biq})$$
$$v_u = (\sum_{z=1}^{m} p_z, \sum_{z=1}^{r} c_z, \sum_{z=1}^{t} w_z, \sum_{z=1}^{q} bi_z)$$
$$v_S = (v_p, v_c, v_w, v_{bi})$$

(5)

We call *vu* the vector of the under-variables *p*, *c*, *w* and *bi*, and *vs* the vector of the vectors of the under-variables' parameters (Fig. 1).

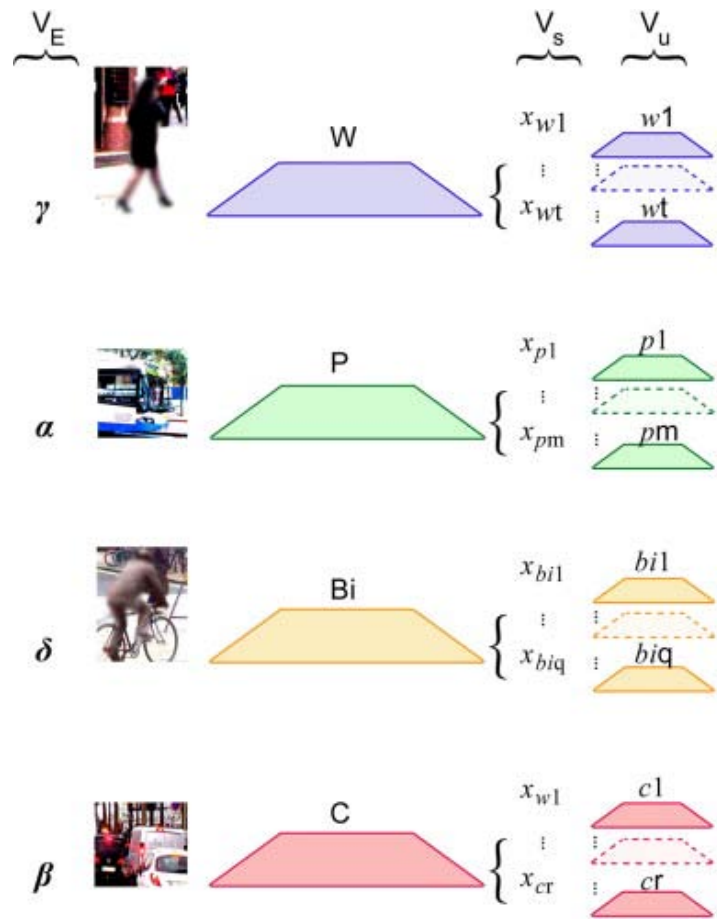

Fig. 1.
Variables, under-variables and parameters of *E*.

The *objective* status of movement possibilities offered by the city is expressed by *v_u*.

The *subjective* status, 'transducer' of the preferences heterogeneity of citizens, is expressed by *v_E*, *v_S* and *ε*, and in the quantification of those under-variables which are not quantitative and objective but qualitative and subjective.

Therefore *v_E*, *v_S* and *ε* transform the Euclidean distance (*d*) into a *Psycho-Economical* Distance, where we think about time in monetary terms.

If we formulate *A* and *E* on aggregate statistics and habits by observing the majority of the citizens rather than each of them, we talk of *Isobenefit Lines*.

When *A* and the full set of parameters of *E* are formulated on *personal* habits and preferences, we pass from *Isobenefit Lines* to *Personal Isobenefit Lines* ( D'Acci, 2013a and D'Acci, 2013b).

For example, a person can give more importance to libraries, parks and pedestrian areas, another to shopping malls and parking, another to hospitals and public transport, etcetera. In the same way, a person could give more value to esthetics and silence, another more about speed and time, another more about costs, etcetera (D'Acci, 2013c). Also, a same person can change her/his needs/preferences along the years, as the daily life of children, teenagers, adults and seniors are different. Again, a same person could change his/her habits when becomes richer or poorer, or get married or divorced, or have babies, or memories, etcetera.

To change needs, habits, preferences means to change the values of the parameters, which in turn means to modify personal *IsoBenefit Orography* (the 3-dimensional visualization of Eq. (4), of the city, which we call also *IsoBenefit Landscape*). Future developments of Eq. (4) may need to consider that amenities do not always have a simply linear addition of their own attractiveness; unless we use directly the contingent attractiveness as the value of punctual benefit ($A_i$), as described in paragraph 9.

## 2.1. Extensions to the basic version

### 2.1.1. Different *E*

The equations of isobenefit lines shown till now, refer to the case of an isotropic area with the same, even if subjective, *E*. We can relax this simplistic version by introducing some extensions to the original model.

In order to treat areas with different *E* we will use the following reasoning and equations illustrated in Fig. 2. For making understanding easier, Fig. 2 uses a linear function between benefit and distance.

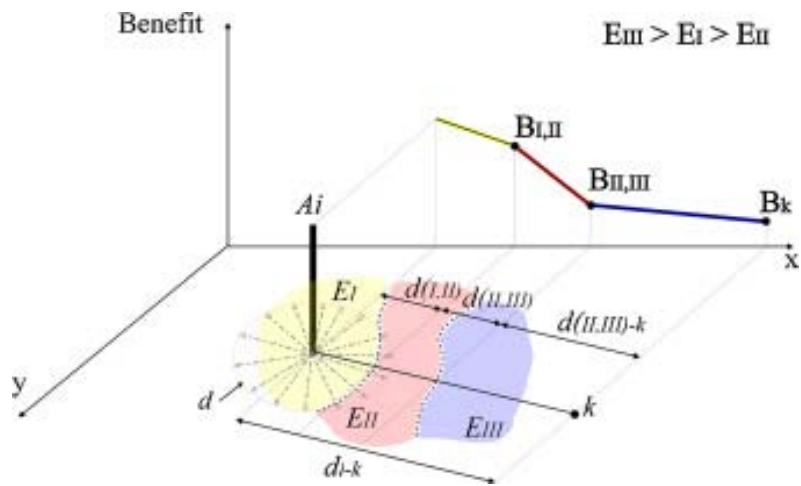

Fig. 2.
"Different *E*" effect on the diffusion of the benefit of an amenity.

Calling $\eta$ the area with a constant $E$, and having:

$$\eta = I \div \psi$$

(6)

$\Psi$ is the number of $\eta$ crossed between $k$ and $A_i$, where the closest to the latter is I, then II, then III, then IV, etcetera till the last $\eta$: $\Psi$.

For calculating the distance between the generic urban point $k$, and the generic attraction $A_i$, we move radially from $A_i$.

For $\eta = I$ we use the usual Eq. (1):

$$B_{i,k} = \frac{A_i}{1 + (d_{i-k}/E_I)}$$

(7)

where $B_{I,II}$ is the benefit received from $A_i$, in the point on the board between I and II ($d_{I,II}$ in Fig. 2), that means $B_{i,k}$ of Eq. (7) when $d_{i-k}$ is $d_{I,II}$:

$$B_{I,II} = \frac{A_i}{1 + (d_{(I,II)}/E_I)}$$

(8)

for $\eta = II$:

$$B_{i,k} = \frac{B_{I,II}}{1 + (d_{(I,II)-k}/\varphi E_{II})}$$

(9)

It is like if we reiterate $B$ into $A_i$, by changing the latter with $B_{I,II}$. We need a corrective coefficient, $\varphi$, in order to avoid the following situation, against the common sense of reality:

we imagine to cross two areas with different $E$ ($E_I$ in area I, adjacent $A_i$, and $E_{II}$ in area II) between $A_i$ and $k$.

If $E_{II} > E_I$, we do not expect to find in any point in area II:

1. a lower benefit than the benefit that the same point will receive by carrying on with $E_I$ rather than $E_{II}$;

2. a higher benefit than the benefit in area I on the boundary with area II ($B_{I,II}$), because, even if the part of the route in area II is much more pleasant, in any case, we will have to pass throughout area I too, and for the same length.

Therefore the points will be inside the blue area of Fig. 3.

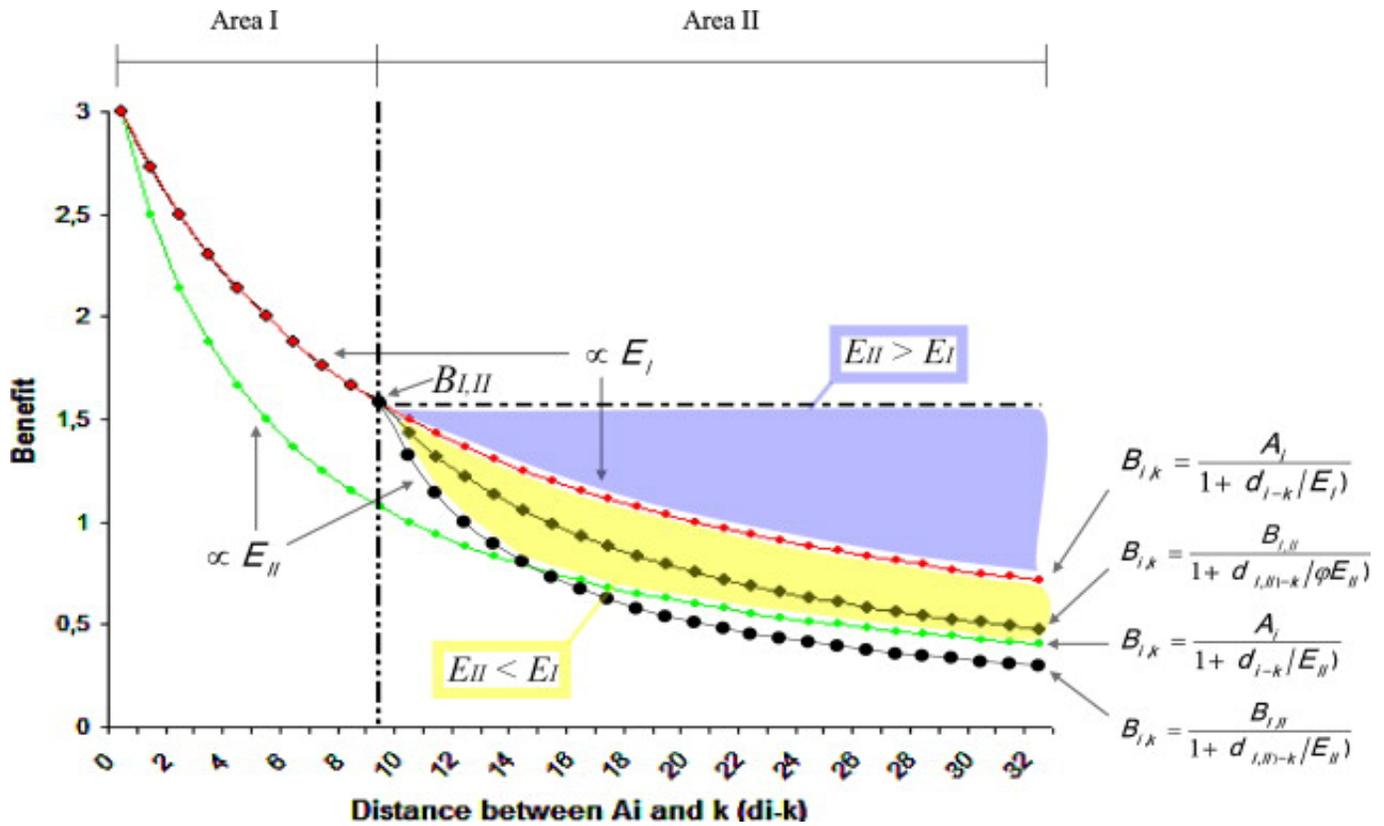

Fig. 3.
The φ influence.

If $E_{II} < E_I$, we do not expect to find in any point in area II:

3. a lower benefit than the benefit that the same point will receive by $A_i$ using $E_{II}$ for all the distance ($d_{i-k}$);

4. a higher benefit than the benefit that the same point will receive by carrying on with $E_I$ rather than $E_{II}$;

Therefore the points will be inside the yellow area of Fig. 3.

Points 1 and 3 are guaranteed by the right value of φ, which is a number major or equal to 1; points 2 and 4 are guaranteed by substituting $A_i$ with $B_{I,II}$ on the equation.
All the above points are against common sense as long as we admit only the perpendicular route between $A_i$ and $k$. However, when later we will consider the more realistic case to choose different routes in order to get $A_i$ from $k$, these two points are not valid anymore.

For η = III:

$$B_{i,k} = \frac{B_{II,III}}{1+(d_{(II,III)-k}/\varphi E_{III})}$$

(10)

And so on; more generally:

$$B_{i,k} = \frac{B_{\eta,\eta-1}}{1+(d_{(\eta,\eta-1)-k}/\varphi E_\eta)}$$

$$\eta = 1 \rightarrow \quad B_{\eta,\eta-1} = A_i; \quad d_{(\eta,\eta-1)-k} = d_{i-k} \quad \varphi = 1$$

(11)

Later, in another paragraph, we will add a finale step which makes the reasoning and the equation more realistic, which will also consider the possibility of citizens to choose the routes. The $B_{i,k}$ to be finally considered is the maximum $B_{i,k}$ among all the possible paths connecting $i$ and $k$. This same final step will regard also the Barriers reasoning and equation.

### 2.1.2. Barriers

The model can also be extended to include barriers effect on the $B$ propagation.
Barriers might be congested car streets, steep climbs, train lines to be crossed by overpass, a very sharp change of $E$ which sign a strong and snap alteration in one or more factors involved in the formation of $E$ (i.e., crime, social segregations, run-down esthetical qualities, etc.).
Calling $E_b$ the $E$ connected to the crossing of barrier $b$ ("how physically and/or psychologically hard it is to cross it"), $B_{i,b}$, the benefit along the barrier (more exactly, just adjacent its opposite side in respect to $A_i$) given by $A_i$, is given by the following:

$$B_{i,b} = \frac{A_i}{1+(d_{i-b}/E_b)}$$

(12)

Therefore, the benefit propagation on the opposite side of the barrier in respect to $A_i$ is, as before, given by:

$$B_{i,k} = \frac{B_{\eta,\eta-1}}{1+(d_{(\eta,\eta-1)-k}/\varphi E_\eta)}$$

$$\eta = 1 \rightarrow \quad B_{\eta,\eta-1} = B_{i,b}; \quad d_{(\eta,\eta-1)-k} = d_{b-k}$$

(13)

where the distances are taken, as before, by aligning $k$ to $A_i$ radially moving from the latter.

Fig. 4 shows an example with linear function to facilitate the reading.

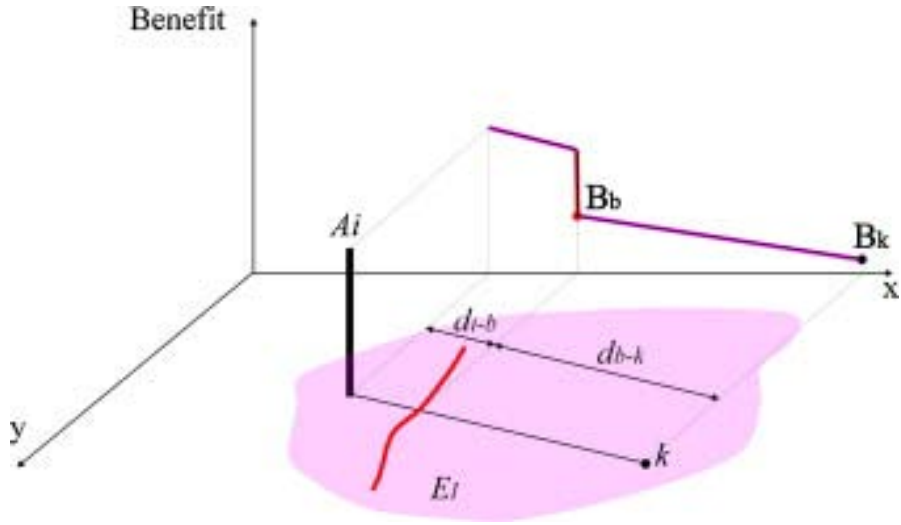

Fig. 4.
"Barrier" effect on the diffusion of the benefit of an amenity.

As for the previous case, later we will add the final step by adding the possibility for the citizen to choose its path to get $A_i$, and therefore the distances will not be anymore radial from $A_i$. In fact the benefit in k given from $A_i$ will be the maximum one received by using all the possible routes. Independently from this addition which allows to decide the rout between k and $A_i$, this passage is automatic when we treat barriers as rivers or train lines which can be crossed only in specific points (bridges or underpasses), or which cannot be crossed at all and therefore been obligated to a different path in order to get $A_i$ from k.

### 2.1.3. Preferential pathways (pedestrian and cycle paths, underground, fast streets, etc.)

In order to consider the effect of pedestrian and cycle paths, underground transport systems, fast public transport lines, fast car streets, etcetera, (we generically call them "paths", or "l" for "line") we extend the model in the following way.

Step 1:
In respect to an $A_i$, we "stop" and reiterate (as done from Eq. (9)) the highest B "touched" by the path received by $A_i$, which we call c and we calculate, strictly along each z point of the path l, the benefit $B_{i,l}$ given by $B_{i,l}$, propagated through $E_l$ and a corrective coefficient $\varphi$:

$$B_{i,z} = \frac{B_{i,l}}{1 + (d_{(i,l)-z}/\varphi E_l)}$$

(14)

where $d_{(i,l)-z}$ is the distance between the point (i,l) which has $B_{i,l}$, and the generic point z along l.

Step 2:
We diffuse $B_{i,z}$ all around, by a diffusion function $f(B_{i,z}, d_{z-k})$ which considers, by $E = E_{k-z}$, again, the psycho-economical distances among k and z:

$$B_{l,k} = f(B_{i,z}, d_{z-k}, E_{k-z})$$

(15)

This equation might have the form of Eq. (14).
We call $B_{l,k,max}$ the maximum benefit in $k$ received by using the path, and it is the benefit in $k$ given by the most convenient $z$ point of $l$ in order to go to $A_i$. We call this point, $z_{max}$. Therefore:

$$B_{l,k} = f(B_{i,z_{max}}, d_{z_{max}-k}, E_{k-z_{max}}) = B_{l,k,max} = B_{z_{max},k}$$

(16)

Step 3:
Finally, we chose as benefit in $k$ received by $A_i$ ($B_{i,k}$), the maximum benefit in $k$ between that one received by using the path (Eq. (16)), and that one received without using it (Eq.(11)).
In a certain way, this final step defines the areas of influence of the path.

## 3. Some examples

We imagine a citizen (who we call Person 1) who uses with the same level of personal preference, indifferently and alternatively (i.e., following the weather, the daily mood and time), public transport, biking, walking and car. This means having a same value of the parameters α, β, γ, δ.
Now imagine two areas within her/his city, area I and area II, whose variables $P$, $C$, $W$ and $B_i$ for Person 1 have different values; for example area I has wonderful pedestrian and cycle paths, perfect public transport services as well as comfortable and not congested streets for cars (i.e., $P$, $C$, $W$ and $B_i$ = 10), while area II does not have pedestrian and cycle paths, and the public transport is not so efficient, while the streets are still excellent for car traffic (i.e., $P = 3$, $C = 10$, $W = 4$ and $B_i = 1$). In this example the points in area I have, for Person 1, $E = 10$, and the points in II have $E = 4.5$.
This means that Person 1 will move much better and more comfortably within area I than area II.

Now we imagine a Person 2 with same personal criteria of Person 1 for judging qualitative under-variables (i.e., esthetical quality of paths, design attractiveness of trams and buses, pleasantness of the background music of the underground transport systems, etc.), and same personal preference among the under-variables relatives importance and evaluation ($v_S$), but with different habits-preferences in ways of moving (different $v_E$). In the same way we imagine Person 3 and Person 4 having different evaluation criteria for reading the qualitative under-variable which are part of $v_u$ and different $v_S$ from Persons 1 and 2. For the same scenario offered from the city, Persons 1, 2, 3 and 4 may associate a different value of $E$ ( Fig. 2).
Fig. 5, Fig. 6, Fig. 7 and Fig. 8 show an example of a sudden alteration of the pleasantness of a path, which may be the case when we have to cross, especially if walking, a congested road or a train line. A similar case may occur for climbs or other kinds of physical obstacle when the person prefers walking or biking throughout the city.

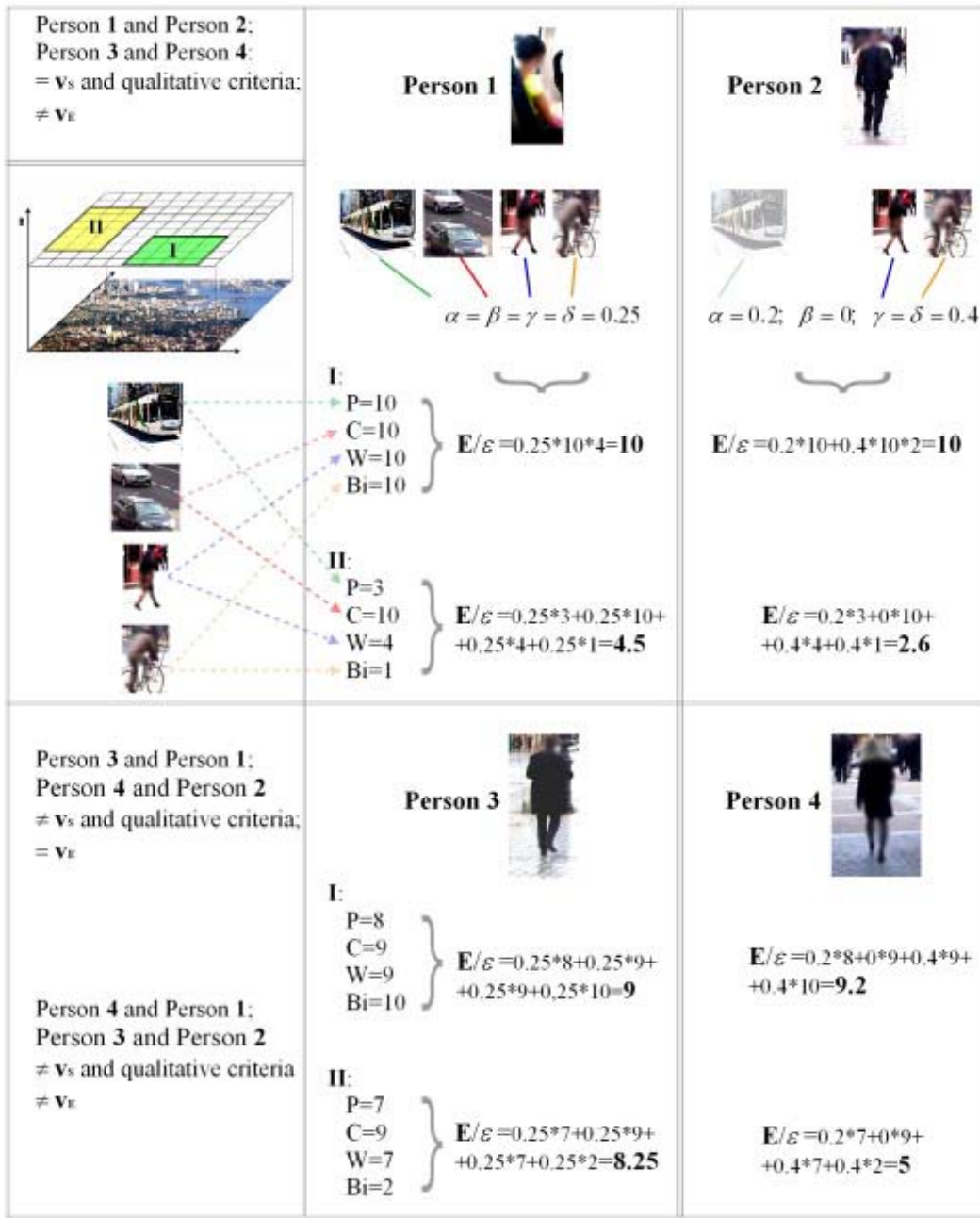

Fig. 5.
Examples of personal evaluation of *E*.

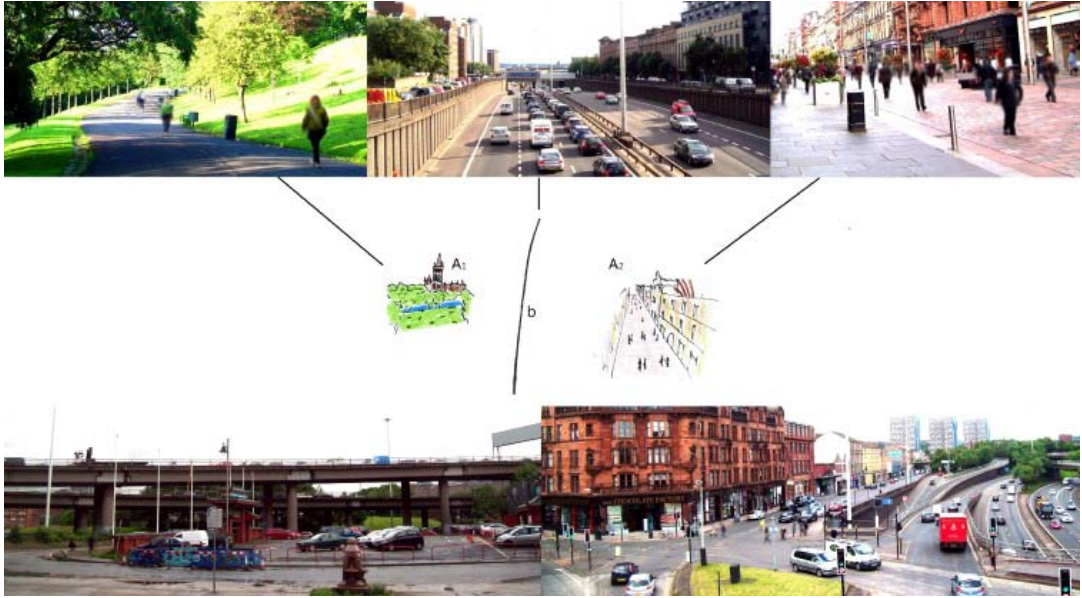

Fig. 6.
Two amenities (*A*1 and *A*2 on the top left and right), and a barrier example (*b*, on the top in the middle and on the bottom).

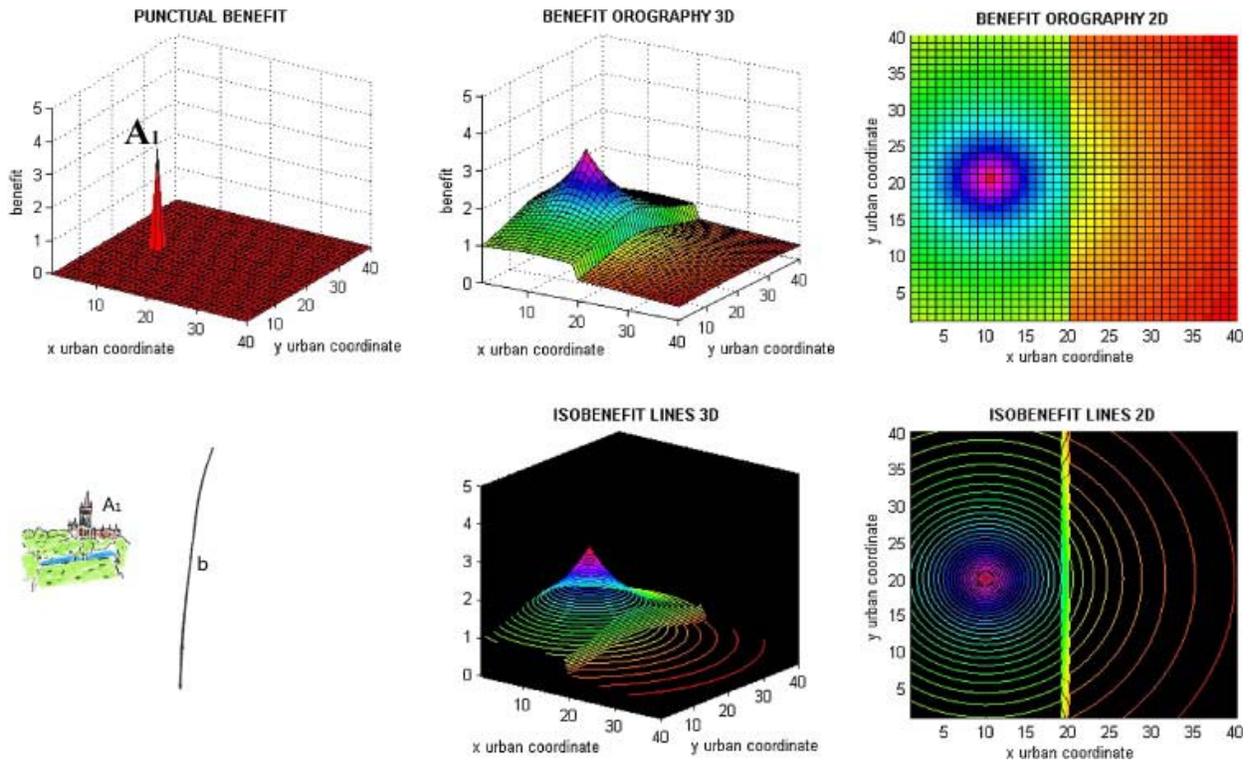

Fig. 7.
Amenity on the left of a barrier.

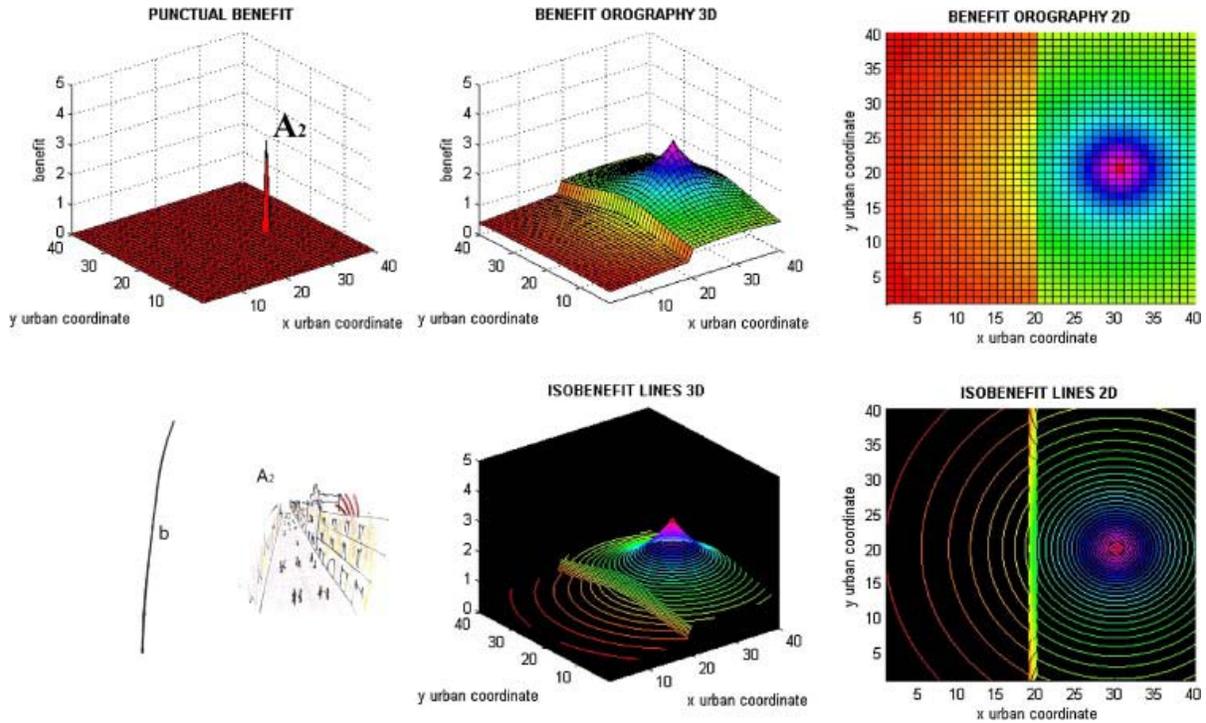

Fig. 8.
Amenity on the right of a barrier.

When we delete the barrier, the benefit in each point of the area increases as well as the uniformity of its distribution (*U*). The latter is estimated by subtracting from one the standard deviation of $B_k$ of every urban point divided by their average value. Therefore *U* is a number less or equal to 1, where 1 indicates the maximum possible uniform distribution. In this example, by removing the barrier, the average benefit increases from 2,0154 to 2,3924, and its uniformity from 0,7820 to 0,8033.

Fig. 11 shows the isobenefit lines effect of a pleasant cycle and/or pedestrian path or an efficient public transport line (Fig. 9). In the latter case distances are not just *phsycologically* shorter, but also *temporally* shorter. Thinking about pleasantness and time in monetary terms, we understand the suffix "psycho-economical distances" ( Fig. 10).

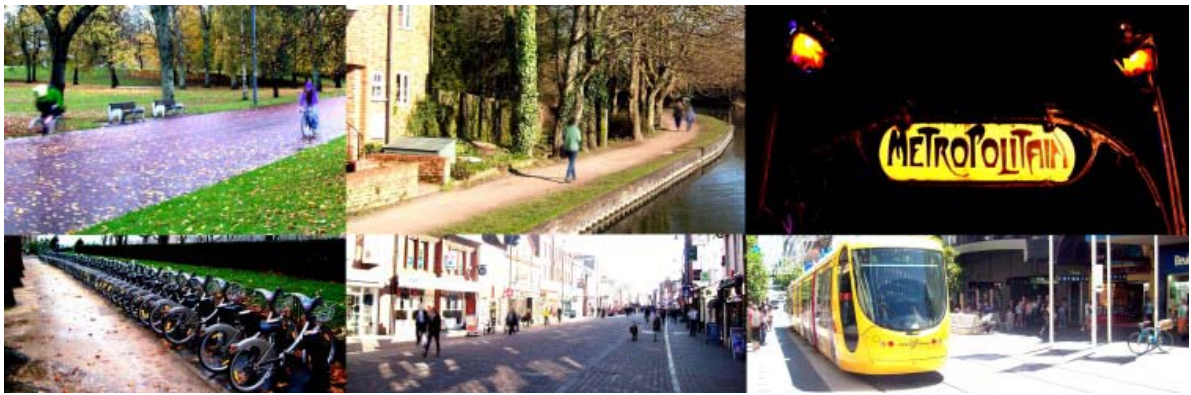

Fig. 9.
Examples of preferential pathways.

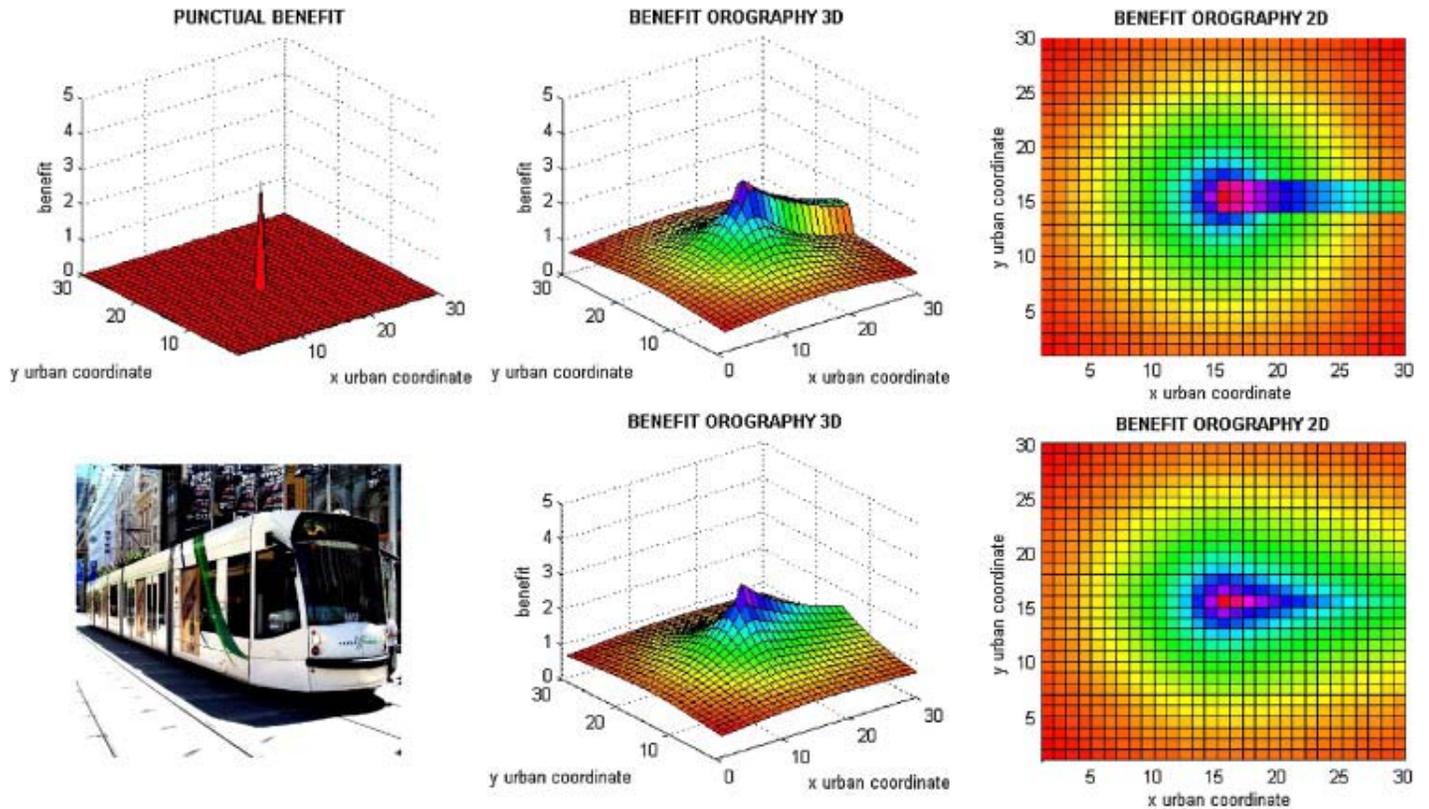

Fig. 10.
Example of preferential pathways with (bottom) and without (top) the diffusion function (Step 2 – Eq. (15))

The top of Fig. 11 does not consider the diffusion effect of Eq. (15), which instead is included at the bottom.

Fig. 11 shows an example of a personal valuation of $E$ for a person who prefers walking and biking for moving throughout the city (in the example of the figure the city is Glasgow). It is a simplified example in which the person uses only two different levels of $E$: $EI$ for the areas on the city where she/he moves (walking and cycling) with a high level of pleasantness; $EII$ for the areas where she/he moves uncomfortably. We remind how the $E$ evaluation may be even exactly the opposite for a person who prefers moving throughout the city by car.
This is just a simplified example to explain the concept, more detailed analysis will have more than just two $E$ across areas within a city.
Fig. 12 shows an isobenefit vision about the difference between an isotropic area (from the $E$ point of view), and an un-isotropic, having a different value of $E$. In the example of Fig. 12 the area on the north of the amenity is less pleasant/efficient from the point of view of $E$, than that one on the south; that means having a different value of $E$ rather than constant. This example is similar to the case of Turin, where the city center benefit distribution is not isotropic because the north area has a general lower quality in respect to the south.

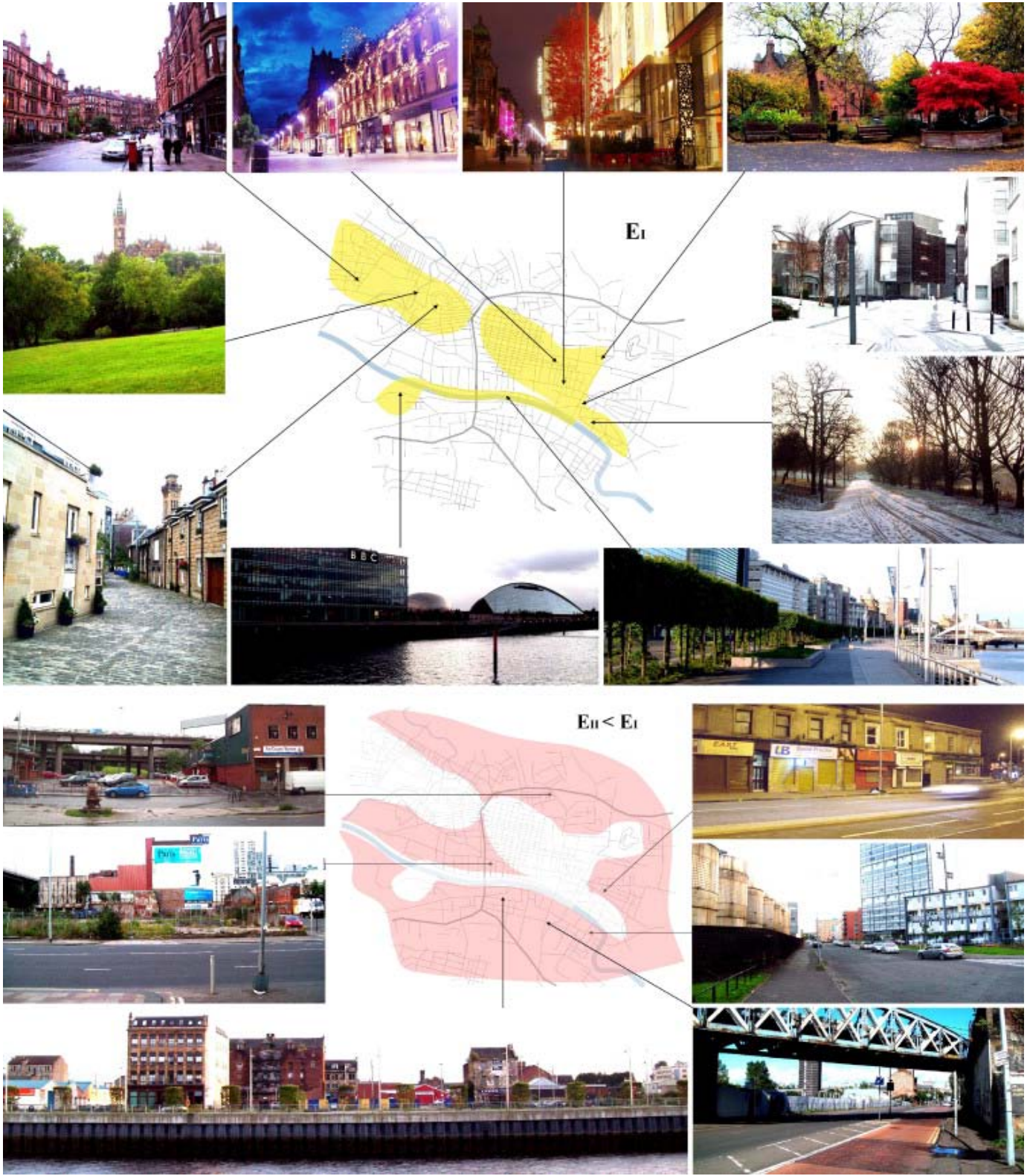

Fig. 11.
Example of a simplified evaluation of *E* for a certain walker citizen profile: areas in yellow (top) have *EI*,
areas in red (bottom) have *EII*, which is minor than *EI*.

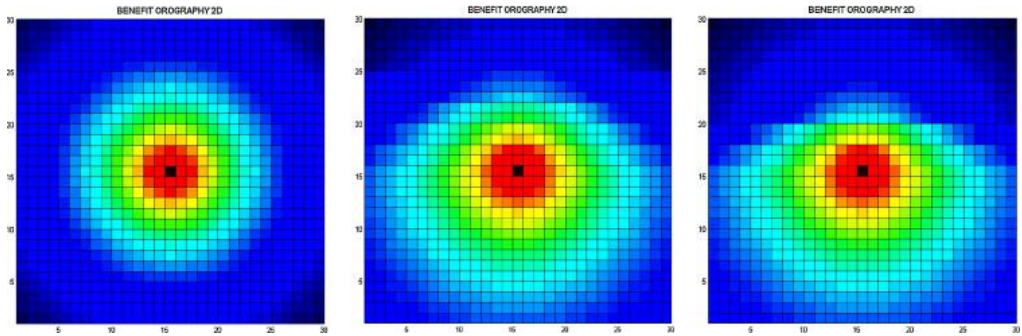

Fig. 12.

Left: example of one amenity in an isotropic area (*E* = constant). Middle: the same amenity but in un-isotropic area with a lower *E* on the north area of the city. Right: same example of the "Middle" but with an even lower *E*.

## 4. Urban planning

Embellishments and improvements due to urban planning, such as pleasant pedestrian/cycle paths, might have a double effect:

1. increasing the attractiveness of the embellished path (mainly, increase *A* of Eq. [(1)](#));
2. reducing (psychologically) distances between points touched directly or within a reasonable distance by the embellished path (increase *E* of Eq. [(1)](#) along the path).

The latter point may occur also not only for linear urban improvements (nice streets and paths), but also for punctual (a square, a corner, a small garden) and fuzzy (a soft but entire improvement of the quality of an area by small diffused urban and architectonical design actions: fuzzy urban quality) urban embellishments.

Concerning the first point, we can have three cases depending on the status before and after the urban improvement ([Fig. 13](#)): a disamenity becomes an amenity (case I); a neutral area becomes an amenity (case II); an amenity improves its attractiveness (case III).

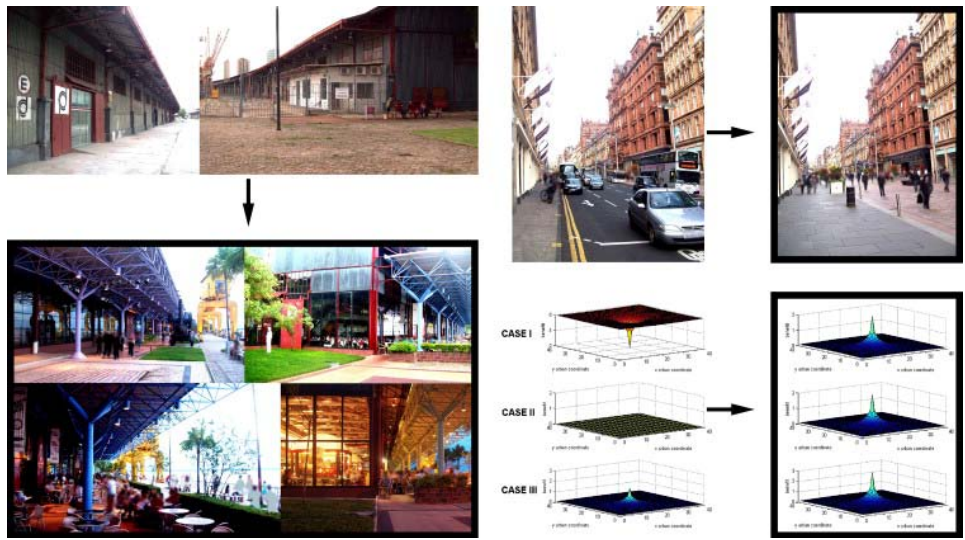

Fig. 13.
Urban transformations and attractiveness ($A_i$) effects.

Fig. 14 proposes a graphical and conceptual summary of the Isobenefit vision of cities.

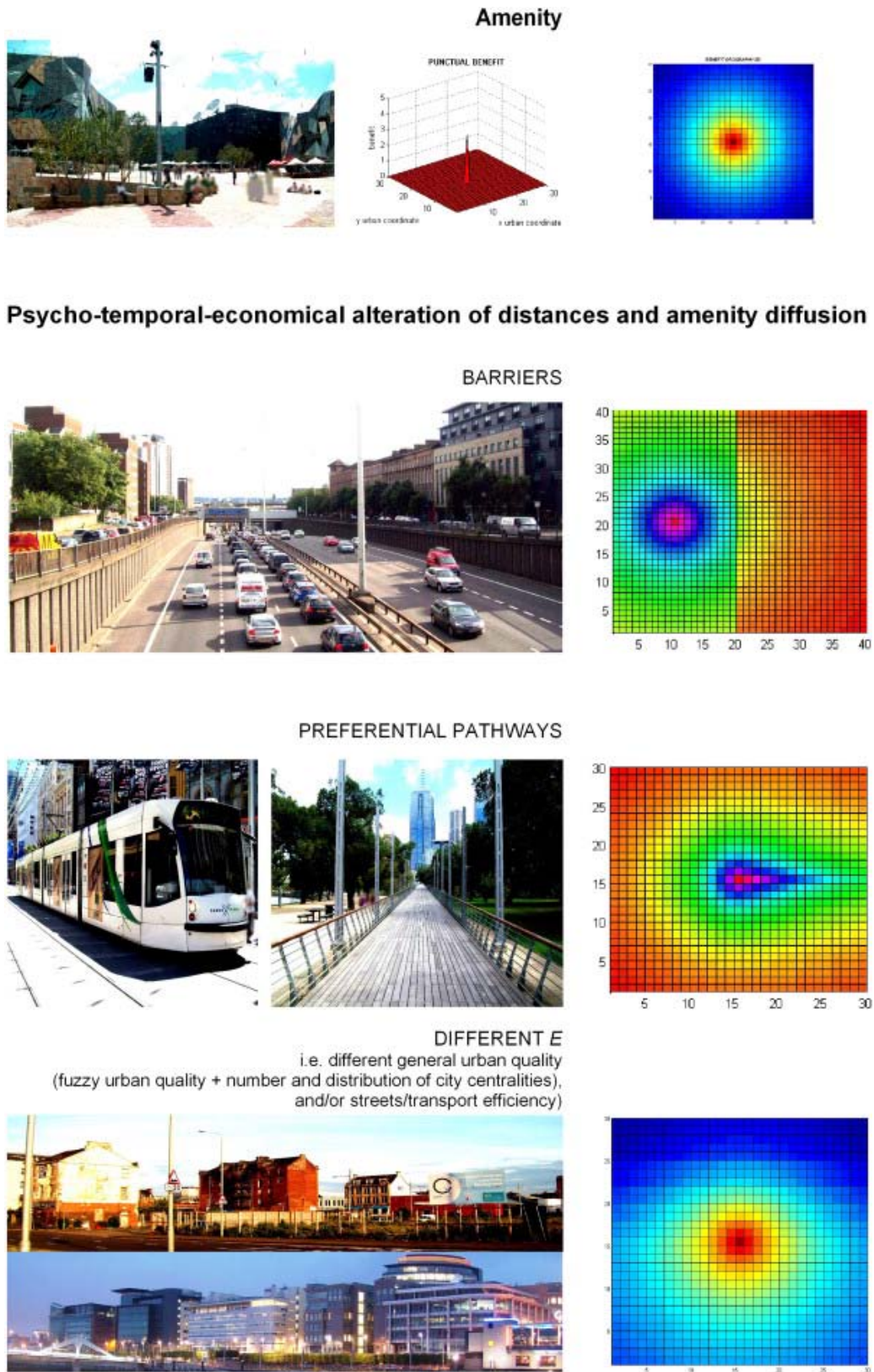

Fig. 14.
Urban isobenefit orography and distances alteration.

## 5. Citizens movement decisions within cities

Psycho-economical distances and personal isobenefit lines may also serve as bases, or parallel tools, to understand citizen movement decisions within cities: where people choose to walk, drive, cycle, and – including as input some fundamental variables such as economics personal budget and work, family and best friends locations – where to decide to live.

Returning to the speech about the diffusion throughout the city of the benefit given by an amenity, and adding the possibility for citizens to choose which route to take toward the amenity, we are now a little step closer to the reality. The benefit in a generic point $k$ given by an amenity $A_i$ is the maximum benefit among the benefits in $k$ given by $A_i$ assuming $d_{i\text{-}k}$ not anymore just the Euclidean distance (even if already translated into psycho-economical distance by $E$) between $A_i$ and $k$, but assuming all the possible routes between them, that means different distances, different $E$ and different spatial scenarios encountered (barriers, preferential paths, personal memories and preferences, etc.).

Calling $r$ the generic route between $A_i$ and $k$; $R_{ik}$ the set of all the possible $r$; $b_r$ the benefit in $k$ given from $A_i$ when the citizen reach it by $r$; $B_r$ the set of all the possible $b_r$. Therefore $B_{i,k}$ is the maximum $b_r$ of $B_r$:

$$B_{i,k} \in B_r | \forall b_r \in B_{ir} \quad B_{i,k} \geq b_r$$

(17)

Fig. 15 shows an example of four different routes, r1, r2, r3 and r4: even if a route is longer, it may be more pleasant and therefore one might prefer it rather than a shorter but less pleasant one.

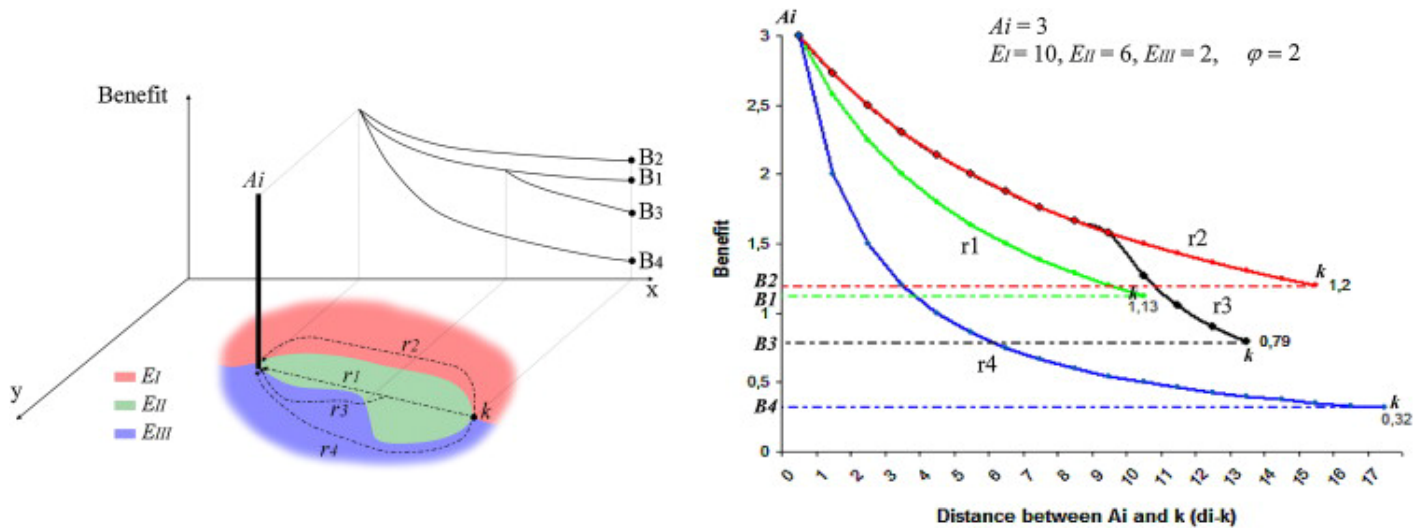

Fig. 15.
Benefit from different routes.

The $B_k$ in the example of Fig. 15 will be B2, mainly the benefit given by $A_i$ through r2.
Summarizing the benefits received in $k$ by all the amenities in the cities, we obtain the benefit in $k$, $B_k$ (Eq. (4)).

## 6. Breaking points of equal attractions

In economy geography, gravitational models were investigated for many decades. Huff (1963) provides the probability (P) of a customer (C) living in a place (i), to travel to a particular facility (j) distant dij, considering all the other n facilities available:

$$P(C_{ij}) = \frac{S_j / d_{ij}}{\sum_{k=1}^{n} S_k / d_{ik}}$$

(18)

By substituting facility with amenity – urban attractions – and customer with citizen, Eq.(6) can be read in another point of view. The same can be done for several gravitational models available. For example, the Isobenefit lines can offer a comparison with the Breaking point of Reill's law of Retail Gravitation (1931). It describes the breaking point of the boundary of equal attraction. Reading Reill's equation from the point of view of urban amenities:

$$Br_{1,2} = \frac{d_{1,2}}{1 + \sqrt{A_1/A_2}}$$

(19)

where $Br$ is the breaking point between the urban amenities 1 and 2, $d_{1,2}$ is their reciprocal distance, and $A$ their attractiveness (punctual benefit). This can be compared with the point, $Br$, shown by the Isobenefit lines in Fig. 17: the minimum value of benefit between the two amenities. It is the point at which a marble placed on the isobenefit surface settles

The breaking point is quite personal: someone (or a same person but at a different stage of her/his life) may prefer going to the closest amenity even if less attractive; that means a negative $\varepsilon$ of Eq. (2).

However, the translation from Euclidean to psycho-economical distances showed how the distances of Eq. (7) can be altered by transport systems and/or physical/psychological barrier, and/or personal preference criteria. When those distance alterations are not symmetrical in relation with the two amenities considered in Eq. (7), the marble slips to one side or the other (Fig. 16).
.

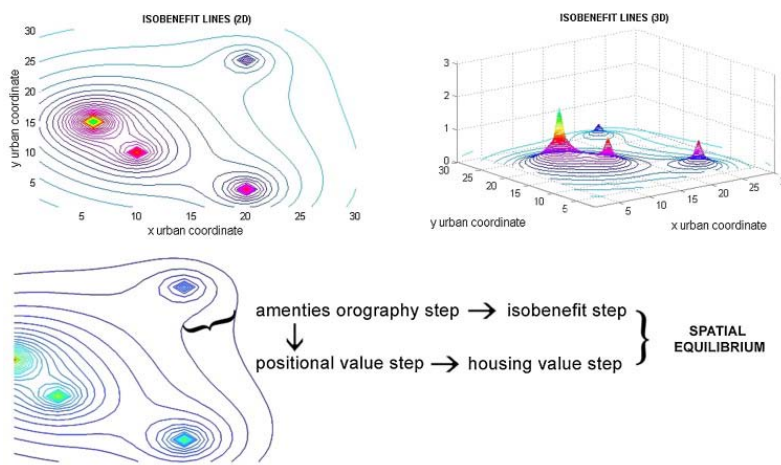

Fig. 17.
Spatial equilibrium within city and isobenefit lines.

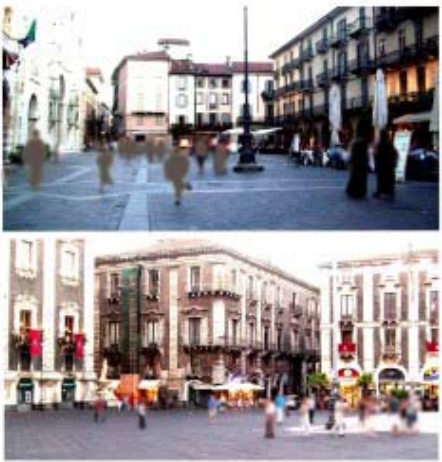
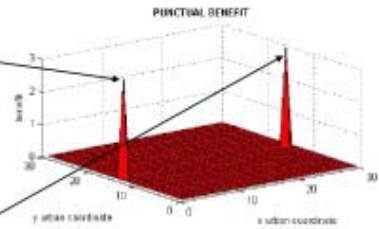
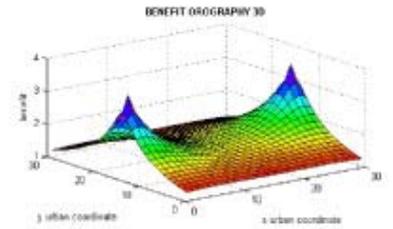
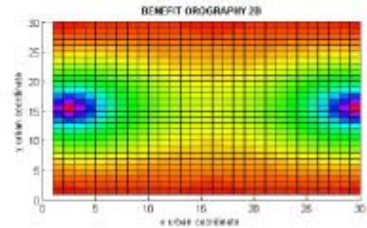
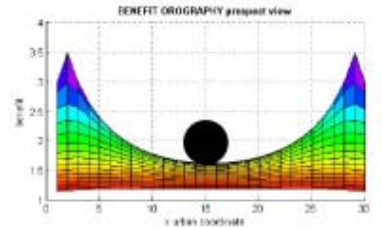
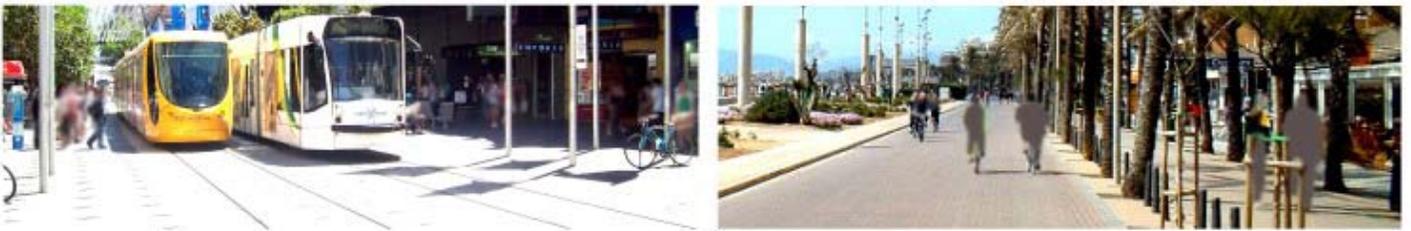
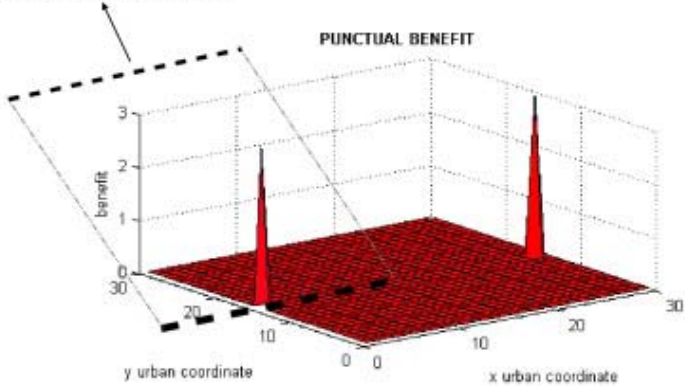
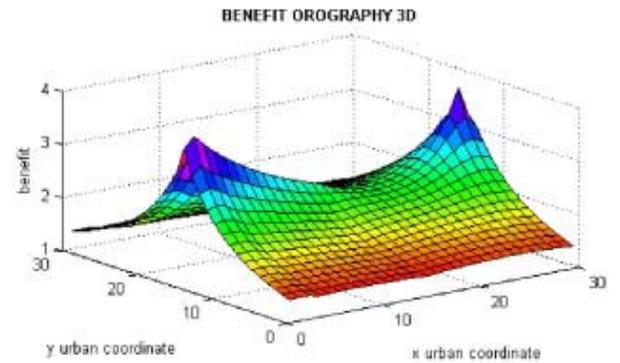
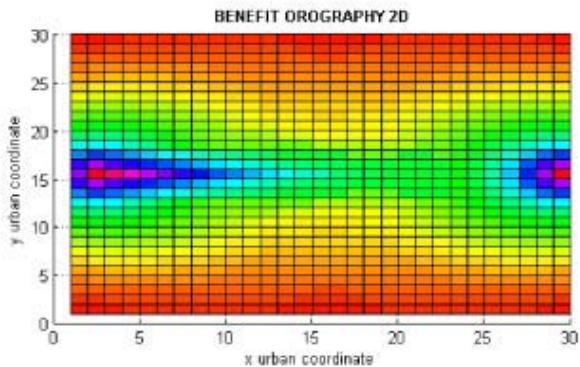
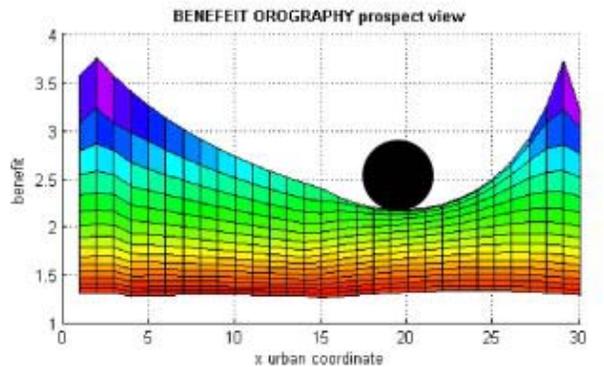

Fig. 16.
Example of breaking points.

## 7. Spatial equilibrium

It is not an equilibrium in terms of time and among people, but in terms of *contingent* situations as for example: I live in a small village just outside the city where I work, therefore I pay *less* for my house but *more* for commuting. The equilibrium resides in this "*less*" and "*more*".

In general simplified approximation, we can think of spatial equilibrium as a compensatory equality throughout the space of the relations among income, prices and amenities.

Spatial equilibrium assumes that these three factors are, under certain conditions, offset against each other when moving from one location to another: high incomes are offset by high prices (housing, cost of living), and/or local negativity (climate, crime, congestion, etc.), and vice versa.

When the study is addressed within a city by comparing the diversity (assumed to be compensatory) between its different urban areas, it is called Spatial Equilibrium Within Cities whose simplest form is the Alonso–Muth–Mills model.

When it is turned to the comparison between different cities, comparing their incomes, costs and amenities, it is called spatial equilibrium across cities whose fundamental model is the Rosen-Roback's.

The isobenefit lines approach offers lines of equilibrium in terms of the benefit that each point receives from the amenities/disamenities present in the city. Calling the positional value the property value given by extrinsic characteristics (not related to the property itself, but to the urban area in which it is located, that means to its amenities/disamenities), its relationship with the isobenefit lines is obvious, and supports the spatial equilibrium assumption according to which a greater real income is offset by something negative, or, in other words, considering the nominal income as a constant, if you choose to live in a very privileged area of the city, real income (here intended as nominal income minus housing costs) decreases because it is inversely proportional to the quality of the area: the greater the area's quality, the greater the positional value, and, therefore, the greater the housing price, and therefore the lesser the real income. Vice versa for a low level of amenities.

Steps among isobenefit lines correspond to steps among urban amenities, that means steps among positional values, that means (ceteris paribus, namely keeping the intrinsic characteristics constant: a same typology, surface, levels, furniture, etc. of the apartment) steps among property values (Fig. 17).

Also in analysis of spatial equilibrium across cities, isobenefit lines may, with due precaution, be a tool of analysis through the comparison of Isobenefit indexes such as the uniformity of the amenities spatial distribution (relative to the urban pleasantness, and/or specific services/amenities one wants to study), their total values, the maximum and minimum value, the centroid coordinates, and so on.

*Isobenefit Lines* are an as easy as powerful tool which transforms cities in tridimensional solids whose shapes depend on the level and spatial distribution of amenities/disamenities and which, for some aspects, offer a more efficient way of comparison and analysis.

Spatial equilibrium is not a *universal* equilibrium, but a personal and contingent equilibrium; it is not an *eternal* or even just a long-term equilibrium, but only a temporary equilibrium, valid just for a specific moment, place, circumstance and, especially, person (see "2.1. Macro and Micro Spatial Equilibrium" in D'Acci, 2013b, p. 6).

## 8. Cities as maths volumes of benefit

Isobenefit Analysis give tri-dimensional volumes whose bases are the city planimetry (x, y coordinates) and the high of each (*x,y*) point is the benefit received from all the amenities in the city through the planimetry characteristics which are included into *E*. The latter, as we saw, involves the Urban Fuzzy Quality (city's beauties at *local level* such as a tree in front of the house, a couple of nice benches, agreeable streetlights, etcetera; is a fuzzy, soft urban quality), efficiency of public transport, congested streets, social contest, crime, personal memories and preferences, and so forth.

The final result is an isobenefit orography of the city (the tri-dimensional visualization of isobenefit lines). If we vertically "close" the isobenefit orography on its sides (the end of the planimetry considered, which in the examples show in this paper are the four sides of the squared planimetry), we obtain volumes – which we call *V* – that we can treat both visually and mathematically in order to be compared. The comparisons might result useful if we want to compare a same city before and after some deep urban transformations, or (with some technical and conceptual devices especially concerning scales of measurements) to compare different cities among themselves.

A few useful indicators, which must be read all together rather than separately, are the following.

We call *M* the matrix of *m* cells we divide the city into, where the generic *k* point, with the height $B_k$, is the central point of the generic cell *K*, and *B'* is the average value of *M*:

$$M' = B' = \frac{1}{m} \sum_{k=1}^{m} B_k$$

(20)

To quantify the uniformity of the spatial distribution of the social benefit we use the coefficient of variation (CV) of *M*: the standard deviation of the $B_k$ of every urban point (*σM*), divided by its average value (*M'*):

$$CV = \sigma M / M' = \sqrt{\frac{\sum_{k=1}^{m} (B_k - B')^2}{m}} \Bigg/ B'$$

(21)

The indicator of uniformity dispersion, (*U*) of the social benefit is:

*U* = 1 - CV

(22)

*U* is a number less or equal to 1, where 1 indicates the maximum uniform distribution. It measures the uniformity of the *orography of the social benefit* resulting from the urban attractions.

If we want to introduce 'non-attractiveness' (busy streets, factories, cemeteries, etc.), we give a negative *A*. If we want to consider also the disamenities, for *U* we should separately consider amenities and disamenities ( Fig. 18).

Fig. 18.
From cities to maths volumes of benefit.

We call the area of the *m* cells: ΔS₁, ΔS₂, ..., ΔSk, ..., ΔSm;
and their central points: (x₁, y₁), (x₂, y₂), ..., (xk, yk), ..., (xm, ym).
Their heights are: f(x₁, y₁), f(x₂, y₂), ..., f(xk,yk), ..., f(xm, ym) = B₁,B₂, ..., Bk, ..., Bm
As shown on the top of Fig. 18, the volume of the rectangular parallelepiped with base area ΔSk and height f(xk,yk) = Bk, is given by the product Bk · ΔSk.
Therefore, the following sum gives an approximated volume of the entire *V*:

$$\sum_{k=1}^{m} f(x_k, y_k) \Delta S_k = \sum_{k=1}^{m} B_k \Delta S_k$$

(23)

The exact volume of *V* is:

$$V = \lim_{m \to +\infty} \sum_{k=1}^{m} f(x_k, y_k) \Delta S_k = \lim_{m \to +\infty} \sum_{k=1}^{m} B_k \Delta S_k$$

(24)

This This limit is called the Riemann sums and is denoted by the double integral of f(x,y) over (in our case) M:

$$\iint_M f(x,y)dS = \lim_{m \to +\infty} \sum_{k=1}^{m} f(x_k, y_k) \Delta S_k = V$$

(25)

where M is the region (planimetry of the city) in the xy-plane and f(x,y) is continuous on M and:

$$f(x,y) \geq 0 \, \forall (x,y) \in M$$

(26)

In the case of negative value of B, the double integral indicates a difference between the volume above M but below f(x,y) and the volume below M but above f(x,y). It is called the *net signed volume*.

We can also calculate double integral over nonrectangular regions and in polar coordinates, however, with exception to the simplest cases, it is not so practical to resolve these double integral forms, therefore it is easier to deduct the total benefit by the sum of each B of all the m points in which we can, or decide to, subdivide the city planimetry.

Borrowing concepts from physical bod and gravitational fields referring to inhomogeneous laminas equilibrium, we can find the *Center of Gravity* of our volume of benefit obtained (V).

"Because the body is composed of many particles, each of which is affected by gravity, the action of the gravitational field on the body consists of a large number of forces distributed over the entire body. However, it is a physical fact that a body subjected to gravity behaves as if the entire force of gravity acts at a single point. This point is called the center of gravity" (Anton, 1992, p. 1120).

A lamina is a sufficiently thin object to be considered as a two-dimensional solid. It is called *homogenous* when having a uniform composition and structure (a constant density, where density is the ratio between mass and area), *inhomogeneous* otherwise.

In our case, the lamina is M and the density is given by B. When the benefit is uniform in each point, M is as a homogeneous "urban lamina", otherwise not:

$B_k$ = const $\forall$ k → U = 1 → homogenous urban lamina;
$B_k \neq$ const $\forall$ k → U < 1 → inhomogeneous urban lamina

The coordinates of the center of gravity of V (*Benefit Center of Gravity*) are:

$$\overline{X} = \frac{\iint_M x \cdot f(x,y)dS}{\iint_M f(x,y)dS}, \quad \overline{Y} = \frac{\iint_M y \cdot f(x,y)dS}{\iint_M f(x,y)dS}$$

(27)

It may result of some interest to compare two "urban laminas": the M assuming as density in each cell/point the benefit on the cell/point, with the M assuming as density the population density. Even if their relative centers of gravity will not give useful information for a comparison (a same center of gravity corresponds to infinite possible different scenarios), rather than the centers of gravity, it may be functional overlapping the two urban laminas as a meter of equality of the social distribution of urban pleasantness.

The urban laminas are the three-dimensional solid (V) pressed into a two-dimensional "solids" where the heights of V become the densities of the laminas.

Therefore, rather than passing through urban laminas comparison, we could also lift up the population density as heights and compare then directly with the benefit volume V.

# 9. Isobenefit lines to read complexity in cities: from locations to networks and flows; from aggregated to disaggregated analysis.

Each of us, when deciding how, when and where to go in our daily citizen lives, take part in the larger phenomena of shaping cities. In turn, the change of cities influences the above decisions.

The change of city structures and of personal preferences/needs imply a change of the isobenefit city landscape which, as we saw, is the map of the benefit given from amenities and which depends from the benefit you receive when you are directly enjoying amenities and from how this benefit flows throughout the city.

According to MIT Technology review, "Isobenefit Lines rewrite rules for understanding city life", and have "the potential to change the way planners think about city design".

The traditional model of the monocentric city is being substituted by more complex ones, where a city has several centralities performing diverse functions.

The isobenefit lines model is intended to cope with this improved complexity, with the shown idea to compute the benefit of a particular location to a resident, taking into account the effect of all the various amenities present in the entire city. Then, it calculates locations of equivalent benefit and connects them with so-called isobenefit lines.

By the *Isobenefit Lines* methodology we see cities as "benefit volumes", subjective for each citizen and continually changeable according to: personal moods/needs/preferences, ways to commute, personal memories and urban transformations. We read the attractiveness of urban points and how these attractions/benefits flow throughout the city, depending on how easy, cheap and agreeable the paths to reach these points are. Doing it for each urban point and for each urban attraction, we obtain the *Isobenefit Orography* of the city, namely a map of its urban attractions and how they flow to any point of the city. This is a "liquid" surface rather than solid, as it varies across time and people. It is in the "liquidness" of the isobenefit orography – due to both top-down and bottom-up phenomena – where resides the complexity of cities, their bottom-up spirit, the dynamicity of equilibriums, locations and of networks, which are all part of an inextricable game.

When there is a consistent majority of citizens with similar *Personal Isobenefit Orography*, in some senses, we can calculate isobenefit lines of the average preference and the peaks (punctual benefit) of the isobenefit orography visualize and quantify the *Potential Urban Centralities*, here intended as the places in a city where the majority of citizens would like to go for enjoying the city's pleasantness in their free time. In some senses the highest punctual benefit indicates where people would like to go most. The combination among punctual and distributed benefit, together with those of the amenities (and disamenities) around, says where people actually go. For this we define platonic (or potential) attractiveness and contingent attractiveness, which, when we have a consistent majority of similar preferences among citizens, become respectively *platonic centralities* and *contingent centralities*. For example, let us think about the following question: "how much do you like the idea of a nice urban park?". Your reply represents the platonic attractiveness which the idea of a nice urban park has on you. Now, depending on where that park is placed in the city, you can say how much you actually like and go to that park in that particular context. This is the contingent attractiveness. In other words, an identical park (platonic attractiveness/platonic centrality), may become (contingent attractiveness/contingent centrality) Central Park in New York or a degraded place of crime and desolation. This is the System Retroactive game (SyR game) between top-down (where urban planners and governments place amenities) and bottom-up (how much inhabitants actually like and use them).

Therefore, this approach also helps the understanding and simulating of urban networks and flows within cities, and introduces us to the crucial game between bottom-up and top-down, indicating citizens behaviors and "natural" evolution as a main ingredient in shaping cities.

It proposes a method to simulate flows from origins and destinations within cities: *where* citizens prefer to go, and *how* they prefer to go there.

Batty (2013) bases *The New Science of Cities* on the understanding of cities by studying relations (networks, flows) between places and spaces rather than the intrinsic attributes of place and space (p. 2).

In other words, we should not focus only, for example, on how beautiful a piazza is, but also on how it is inserted within the context of the city: namely how it is connected with the rest of the city, with the centralities, attractions, services, amenities, residences, public transport, streets, promenades, shops, etcetera.

Cities are not a mere sum of isolated objects, but physical manifestations of interactions. Any location we look at in a city is created by relationships through locations: "locations are places that anchor interactions", are "patterns of interactions acting as the glue that holds populations together through flows of material, people, and information", and are "the nodes that define the points where processes of interaction begin and end" (Batty, 2013, p. 8).
Rather than studying locations of things in cities, we should conceive them as networks, relationships and flows.

We must look at cities as "constellations of interactions, communications, relations, flows, and networks rather than as locations", and Batty argues that "location is, in effect, a synthesis of interactions: indeed, this concept lies at the basis of our new science" (Batty, 2013, p. 13).
Flows depend more on activity in locations (size of the nodes which represent the origins and destinations of the flows), on *what* is being connected, *what* there is in locations; while networks depend more on *how* locations are connected (in a way the term $E$ in Eq.(1)).
Therefore, from a predicting point of view, by using multi-agents models we can proceed as the following: knowing (by, i.e., stated preferences) the personal potential/platonic values of each $A$, of Eq. (1), throughout the planimetry of the city, and the personal habits, preferences and criteria which weigh $E$ in Eq. (1) for each citizen (or by dividing the population in several typologies), we may simulate the probability for people to actually go to locations and how they go there (by which paths).
The above will be influenced from the network generated by the relative positions of each location (the urban coordinate of each amenity $i$), their appreciation from each citizen (the *platonic* value of $A$), and how locations are reachable ($E$).
By the above simulation we obtain the contingent attractions which are defined by the emergent flows generated by the combination of the network and personal preferences.

From these contingent attractions we obtain the *contingent* value of $A$ which is finally used for quantifying and visualizing the map of attractions and how they are spread throughout the city (Isobenefit Orography).
We may also simply use the isobenefit lines approach based directly on contingent values of $A$ as a picture of the state of the art of a city in a specific moment.
Almost for any location of any type one can see it is a product or addiction or synthesis of a variety of flows, and these flows are from other positions in the city and they result from networks.

Networks connect locations and the flows among locations, throughout networks, sustain different locations.

In simple words: (1) people go there, *therefore* an attractive square is built; or (2) an attractive square is built there, *therefore* people go there?
The attraction of a location clearly depends on the number of people who go there; but people go there because it is attractive as well.

It is a positive feedback, a two way process.

We cannot say at any point in time that clearly the location is attractive because of *what* is there (even if it could be *more* attractive because of what is there), as it is also because people *want* to go there; and people may want to go (flow) there because that location is close to, or in between, several other locations (network), and/or very well accessible (network), but people may also want to go there purely because the location is attractive per se.
The personal isobenefit lines approach also aims to fit with this need to simulate and predicting networks and flows according to change in locations and urban structure (new streets appear, or new public transport lines, or a square becomes pedestrian, or a new shopping center is built, and so on) and to change in preferences (citizens habits and preferences).

In fact they may be used as a platform on which multi-agents models can run.

Personal isobenefit lines approach also underlines the retroactive games between locations and people by suggesting to see cities as chessboards: "parks, beautiful squares, attractive pedestrian streets, intriguing shopping areas, are like pieces of a chessboard; the potential force of the latter once placed on the chessboard, depends on their positions (absolute and relative), and their use from the players, and is transformed into contingent force, which constantly change at each step of the game.

In a city the pieces are amenities; their potential force are their potential attractiveness; the players are thousands or millions (citizens, investors, governments); the contingent force are the contingent attractiveness, which means the actual use and enjoyment of amenities and urban areas" (D'Acci, 2013b, pp. 15, 16). In other words, a same location (a piece of the chessboard; or a pedestrian square, a shop center, a pleasant park, etc., in a city) once located in the planimetry of the city (the chessboard) changes the current network and flows, and this may happen at both the local neighboured scale and at the entire urban scale, and its level of success (the contingent force) depends from how the location is inserted in the network and flows at both local and urban level.

The isobenefit lines approach used as a platform for multi-agents models, can simulate flows for different scenarios by manipulating locations, networks and personal preferences.

## 10. Some conclusions

Our world view is, consciously or not, constantly shifted between two kind of sights: scientific and humanistic. According to Snow's thesis, "the breakdown of communication between the 'two cultures' is a major hindrance to solving the world's problems" (Portugali, 2011, p. 10). This separation is observable also in urban study approaches: scholars who, by scientific methods tempt to develop a science of cities, and scholars who approach cities with a more humanistic philosophy. During the first part of the 20th century, we see both sides developing in parallel such as the humanistic perspective of Mumford and the quantitative perspective of Christaller, Losh, Reilly and others. In the 1950s a *quantitative revolution* happened, which has "strongly criticized and even de-legitimized the scientific validity of what they have referred to as descriptive approaches" (Portugali, 2011, p. 10). As history teaches us, we live in a sort of perpetual cycle of revolutions and recalls, so that in the early 1970s urban social theories, through Structuralist Marxist and phenomenological idealistic perspectives, strongly criticized the positivistic-quantitative approach.

The last couple of decades have seen both: social theory urban approaches adopting postmodern, poststructuralist and deconstruction philosophies, and quantitative spatial sciences.

However, as Toynbee poetically synthesizes the western democratic ideal to attempt to reconcile two spirits almost diametrically opposed ("The spirit of nationality is a sour ferment of the new wine of democracy in the old bottles of tribalism", Toynbee, 1972, p. 34), history also teaches us the advantages in borrowing the best of each thought along timelines and cultures.

In the discussion of this paper we have found elements of both positivistic-quantitative language (i.e., by the similarity between isobenefit lines and space-syntax, gravitational, rental, spatial and interactional models), and humanistic social language (in the personal readings of preferences, needs, and pleasantness).

From one side, close to the current dominant bottom-up, decentralized vision of phenomena as complex system, as well as to the postmodern urbanism philosophy, the personification of isobenefit lines looks at "society as highly connected but irredeemably plural and contradictory" (Lynch, 1984, p. 46). From another side, close to the top-down, centralized vision, as well as the Post liberal, Modern urbanism believe, isobenefit lines and their variants, aim to understand (bottom-up) agents behaviors – often driven by reductive and short-term views – in order to suggest (top-down) planning which – in comprehensive, long-term view – follow them (when the overall effects of the agents action is positive) or impede them (when they have negative effects such as pollution, congestion, crime and segregation, waste of space, material and energy, unequal enjoyment of city beauties, no green and proper agreeable and liveable public spaces for everyone in cities, etc.).

Theory "must speak to purposes, and not about inevitable forces" (Lynch, 1984, p. 41). However, these purposes (top-down), if they do not want to continue "the failure of an entire discipline, which originated at the end of the nineteenth century around ideas of top-down control" (Pagliardini, Porta, & Salingaros, 2010, p. 331), should first understand, and therefore feel, the bottom-up processes which dry on the basis this complex whole which a city is.

Isobenefit lines approach, in this light, may be looked at like a *Modern Postmodernism*approach, mixing the positive contributes of both (see Isobenefit Urbanism in D'Acci, 2013b, pp. 6, 7):

1. understanding pluralism, genius loci, local values, multi-agents preferences and behaviors;
2. driving cities with long-term and collective wellbeing views.

When urban planners have to manage small towns, most of the intuitions and mental visualization of the urban transformations effects might be done by mind. On the contrary, when the objects are megacities with millions of inhabitants, things become much more complex and Isobenefit lines Analysis show their best usefulness.

By adding the subjective measures, this paper show how, even under an identical objective scenario different people may have different visions. The technical tools of this paper enable us to visualize these different visions.

Each equation proposed is weighted by subjective measures, therefore they all add crucial subjective measures, which is the entire aim of this paper.

A model is a model and not the reality: it translates, in some way, something real. In this case this "way" are numbers, which in turn are intelligibly visualized. Whether we like it or not, all the models turn something (behaviors, actions, facts, preferences, etc.…) into "numbers". In our case, these "numbers" are directly given by people (or by their behaviors: where they go, why and how) expressing their own preferences.

The technical distinction of this paper is the large pattern of parameters offered in order to read subjective criteria. It transformed aggregative preferences and habits to personal ones, as well as isotropic spaces into a more realistic anisotropic space.

Any model is a simplistic version of reality, especially in social sciences. Any model is a simplification of the reality rather than the reality itself; the important point is that it should be a *simplified* mirror of reality instead of a *distorted* one. It should reflect a *simplified*image rather than a *different* one. This paper showed a, necessarily, simplified reality and all the details proposed help the entire approach to be less simplistic.

The innovation of the isobenefit lines approach is to have a formal tool which enables us to think, read, visualize and quantify the complexity and variety intrinsic in our cities and in our personal views of them. We can use flow lines as a platform on which we place agents in order to simulate flows from origins and destinations within cities: where citizens prefer to go, and how they prefer to get there.

We also saw how isobenefit lines offer a tool to identify, visualize and quantify urban centralities, intended here not necessarily as the most *accessible* points in a city and/or the areas where the majority of people must go (for necessity rather than for choice), but as the most *enjoyably* urban places which *please* the ordinary citizen. Urban centralities considered here are not the places where the majority of citizens go because they must, but also because they want to.

Therefore centralities in this sense may certainly involve accessibility, even though without requiring it as a necessary condition to be centralities.

In other terms, differently to current spatial analysis methods, isobenefit lines do not say: these are the most accessible urban points *therefore* they are the urban centralities and where people go most. They say: these are the most agreeable urban points (punctual benefit), and the most reachable in term of cost, time and beauty, or appeal (distributed benefit), *therefore* they are the urban centralities and where people go most, when they chose where they *like* to go. People do not necessarily go in the most accessible points, but they go where they need and want to go, and, again, they flow through paths they need or they choose to pass through. If the most accessible points are, or become, also the most pleasant or where most services and job locations are, this is another fact.

If we use isobenefit lines under the criteria of services, or job locations, we obtain a map and a platform to use to, respectively, visualize and simulate networks and flows connected to the question: where people *need* to go and to pass through (jobs, schools, supermarkets, hospitals, residences …). If we use isobenefit lines under the criteria of pleasant places, we refer to the question: where people *want* to go and pass through.

For the latter reason, they also introduced the "likeability" of places and paths: in addition to the usual parameters currently used (i.e., Space Syntax and Urban Network Analysis), which weight distances in term of physical distance, cost, time or mental easiness representations (number of times a path turns or changes angle), psycho-economical distances used in the isobenefit lines proposed here, also consider how a place and a path *pleases* us.

This pleasure to pass through, or to stay in, an agreeable area also has an underground and an inertia effect (D'Acci, 2013c) which contributes to delight our lives. The final purpose of the science of cities and urban design is to understand cities and make them efficient and attractive to please our lives (D'Acci, 2013a and D'Acci, 2011); "by planning good cities we could increase their attractiveness; not just because in the city one can have a higher income or education, but because one lives in a beautiful city" (D'Acci, 2013a, p. 543).

"Cities are the largest and probably among the most complex networks created by human beings" (Blanchard & Volchenkov, 2009, p. viii). Inside this complexity, beyond practical uses of isobenefit lines Analysis – from urban planners, sociologists, economists, designers, environmental psychologists, real estate agents, multicriteria urban decision makers, morphogenesis studios which might find them useful – this discussion will have served its purpose if it helps a comprehensive view of the "two cultures" without the need to mutually exclude themselves.

We should use *science* to understand evolutionary trajectories in urban phenomena from what concerns objective factors, as well as to deduce universal rules common among cities – such as the scaling laws, power laws, self-similarity, fractal evolutions and structures; we should use *art* to move us, to affect us.

Cities are physical manifestations of our behaviours, chronologically stratified in history, economically localized and shaped, that can be read by mathematical laws, translated into numbers even each keeping their own untranslatable unique art and genius loci. As for a work of art, what we expect in a city is a personal element: each city reveals original, unique elements; each of them is *special* and, even more, is *differently* special for each of us: each city is viewed from as many different private visions as there are persons experiencing them. Personal isobenefit lines visualize by numbers and volumes these infinite private intimate visions.